\documentclass[11pt]{article}
\usepackage{BSMvietnam,epsfig,amsmath,amssymb}

\def\be{\begin{equation}}
\def\ee{\end{equation}}
\def\bea{\begin{eqnarray}}
\def\eea{\end{eqnarray}}

\begin{document}
\vspace*{4cm}
\title{SYMMETRIES IN MULTI-HIGGS-DOUBLET MODELS}

\author{\underline{I. P. IVANOV}$^{1,2}$, E. VDOVIN$^{2}$}

\address{$^1$ IFPA, Universit\'{e} de Li\`{e}ge, All\'{e}e du 6 Ao\^{u}t 17, b\^{a}timent B5a, 4000 Li\`{e}ge, Belgium\\
$^2$ Sobolev Institute of Mathematics, Koptyug avenue 4, 630090, Novosibirsk, Russia}

\maketitle\abstracts{
We report the recent progress in understanding of symmetries which can be implemented in the scalar sector
of electroweak symmetry breaking models with several Higgs doublets.
In particular we present the list of finite reparametrization symmetry groups which can appear in the three-Higgs-doublet models.}

\section{Symmetries in extended Higgs models}

The nature of the electroweak symmetry breaking remains one of the hottest
issues in high-energy physics. 
Properties of the recently discovered \cite{discovery} Higgs-like resonance at 126 GeV
show intriguing deviations from the Standard Model (SM) expectations,
which might be a hint that a non-minimal Higgs mechanism is at work. 
On theory side, many variants of non-minimal Higgs mechanism have been proposed. 
One simple and phenomenologically attractive class of models involves 
several Higgs doublets with identical quantum numbers.
The scalar potential in these $N$-Higgs-doublet models (NHDM) is often assumed
to be symmetric under a group of unitary (Higgs-family) or anti-unitary (generalized $CP$)
transformations acting in the space of doublets.
These symmetries play a pivotal role in the phenomenology of the model, 
both in the scalar and in the fermionic sectors \cite{flavour},
and they often bear interesting astrophysical consequences.

Focusing on the scalar sector,
it is very desirable to know which symmetry groups can be incorporated 
in a model with several Higgs doublets, and how they affect the phenomenology.
In the two-Higgs-doublet model \cite{TDLee} (2HDM), this question has been answered several years ago
\cite{2HDMsymmetries}, see also a recent review \cite{review2HDM}.
Beyond two doublets, the problem remains open. 
Althoug many variants of NHDM based on several finite groups have been studied \cite{variants}, 
no comprehensive list of allowed symmetry groups was known.

In this talk we report a significant progress in this problem.
We found \cite{abelianNHDM} a strategy to classify all abelian groups which can appear as symmetry groups
of the scalar sector in NHDM with arbitrary $N$.
In the case of the three-Higgs-doublet model (3HDM), 
we moved further and listed all allowed finite symmetry groups \cite{finite3HDMshort,finite3HDMlong}.
Our work opens up the way to systematic study of all symmetry-related issues
in 3HDM.

\section{Classification of finite symmetry groups in 3HDM}

\subsection{General strategy}

The most general renormalizable gauge-invariant scalar potential of 3HDM can be written as
\be
\label{V:tensorial}
V = Y_{ij}(\phi^\dagger_i \phi_j) + Z_{ijkl}(\phi^\dagger_i \phi_j)(\phi^\dagger_k \phi_l)\,,
\ee
where all indices run from 1 to 3. We are interested in finite groups $G$ of transformations of doublets $\phi_i$ 
which can be the automorphism groups of some potentials. Such transformations can be either unitary (Higgs-family) transformations
$U$ from the group $PSU(3)$ or antiunitary (generalized-$CP$) transformations of the form $U\cdot CP$. 
The necessity of working with $PSU(3)$
instead of the more common $SU(3)$ in explained in \cite{abelianNHDM}.

Our strategy of finding all finite symmetry groups contains four steps:
\begin{itemize}
\item
find all abelian finite groups which can be subgroups of $G$,
\item
using results from pure finite group theory, establish the general structure of $G$,
\item
after the search is reduced to a finite and small number of possibilities,
check them one by one and see which can be realized in 3HDM,
\item
after all groups of Higgs-family transformations are found, check which of them can be 
additionally enlarged by a generalized-$CP$ transformation.
\end{itemize}

\subsection{Finite abelian symmetry groups in 3HDM}

We developed a machinery \cite{abelianNHDM} which classifies all abelian symmetry groups
in NHDM with arbitrary $N$. In particular, for $N = 3$, we have the full list of such groups
which can be imposed in the scalar sector. They are
\be
\mathbb{Z}_2\,, \quad \mathbb{Z}_3\,, \quad \mathbb{Z}_4\,, \quad \mathbb{Z}_2\times \mathbb{Z}_2\,,\quad 
\mathbb{Z}_3 \times \mathbb{Z}_3\,.\label{abelian} 
\ee
We proved \cite{abelianNHDM} that 
trying to impose any other abelian Higgs-family symmetry group on the 3HDM potential 
unavoidably makes it symmetric under a continuous group (which we disregard by construction).

\subsection{General structure of finite symmetry groups in 3HDM}

Suppose that the potential is invariant under a finite (non-abelian) group $G$ of Higgs-family transformations.
All abelian subgroups of $G$ must be from the list (\ref{abelian}).
By Chauchy's theorem, the order of the group can have only two prime
divisors: $|G| = 2^a 3^b$. Then according to Burnside's $p^aq^b$-theorem, the group $G$ is {\em solvable}.
Solvability implies that $G$ contains a normal abelian subgroup, which belongs, of course,
to the list (\ref{abelian}). 

Let us denote by $A$ this normal abelian subgroup of $G$, $A \lhd G$.
Obviously, $A \subseteq C_G(A)$, the centralizer of $A$ in $G$ 
(all elements $g\in G$ which commute with all $a \in A$).
It turns out \cite{finite3HDMlong} that this $A$ can be chosen in such a way that it coincides with its own
centralizer in $G$ (that is, it is self-centralizing): $A = C_G(A)$.
This means that elements $g \in G$, $g \not \in A$, cannot commute with all elements of $A$.
Therefore, they induce automorphisms on $A$:  $g^{-1}ag \in A$ for any $a \in A$,
and these automorphisms are non-trivial. Even more, 
if $g_1$ and $g_2$ induce the same automorphism on $A$, 
then $g_1$ and $g_2$ belong to the same coset of $A$ in $G$.
Therefore, the homomorphism $f: G/A \to Aut(A)$, 
where $Aut(A)$ is the group of automorphisms on $A$, is {\em injective}.
We conclude that
\be
G/A = K\,,\quad K \subseteq Aut(A)\,.\label{GA}
\ee
This result proves that $G$ cannot be too large,
and it also shows that $G$ can be constructed as an extension of $A$ by a subgroup of $Aut(A)$.
As it will turn out that in all the cases below these extensions are split, we finally find 
$G = A \rtimes K$. 

\subsection{Explicit construction of possible symmetry groups}

The classification problem is then reduced to the following task: 
pick up a group $A$ from the list (\ref{abelian}), calculate its automorphism group
$Aut(A)$, select a subgroup $K \subseteq Aut(A)$, construct all possible non-abelian extensions
of $A$ by $K$, and finally check if this construction fits inside $PSU(3)$.

If $A = \mathbb{Z}_2$, then $Aut(A)$ is trivial, and there is no non-abelian extension.
If $A = \mathbb{Z}_3$, then $Aut(A)=\mathbb{Z}_2$, and the only non-abelian extension $\mathbb{Z}_3 \rtimes \mathbb{Z}_2$
is the group $D_6$, the symmetry group of the equilateral triangle.
One can indeed construct a potential symmetric under this group and --- which is equally important ---
not symmetric under any other transformation. 
If $A = \mathbb{Z}_4$, then $Aut(A)=\mathbb{Z}_2$, and there exist two non-abelian extensions: $D_8$ (symmetries of the square)
and $Q_8$ (quaternion group).
However when constructing these groups explicitly, we find that any $Q_8$-symmetric potential
becomes automatically symmetric under a continuous group.
$Q_8$ is then disregarded, and the only realizable finite group in this case is $D_8$.

This analysis becomes a bit more tricky in the case of $\mathbb{Z}_2\times \mathbb{Z}_2$ and, especially, $\mathbb{Z}_3\times \mathbb{Z}_3$.
Skipping the details, which can be found in \cite{finite3HDMshort,finite3HDMlong},
we give the final list of finite groups realizable as Higgs-family symmetry groups in 3HDM:
\begin{eqnarray}
&&\mathbb{Z}_2\,, \quad \mathbb{Z}_3\,, \quad \mathbb{Z}_4\,, \quad \mathbb{Z}_2\times \mathbb{Z}_2\,,\quad
D_6\,,\quad D_8\,, \quad T \simeq A_4\,,\quad O \simeq S_4\,,\label{list}\\ 
&&(\mathbb{Z}_3 \times \mathbb{Z}_3)\rtimes \mathbb{Z}_2\simeq \Delta(54)/\mathbb{Z}_3\,, \quad 
\ (\mathbb{Z}_3 \times \mathbb{Z}_3)\rtimes \mathbb{Z}_4\simeq \Sigma(36)\,.\nonumber
\end{eqnarray}
This list is complete:
trying to impose any other finite Higgs-family symmetry group on the 3HDM potential
will unavoidably lead to a potential symmetric under a continuous group.

Note that we also gave examples of the potentials for each of these groups.
Just to provide an illustration, here is a $\Sigma(36)$-symmetric potential:
\bea
V & = &  - m^2 \left(\phi_1^\dagger \phi_1+ \phi_2^\dagger \phi_2+\phi_3^\dagger \phi_3\right)
+ \lambda_0\left(\phi_1^\dagger \phi_1+ \phi_2^\dagger \phi_2+\phi_3^\dagger \phi_3\right)^2\nonumber\\
&& + \lambda 
\left(|\phi_1^\dagger \phi_2 - \phi_2^\dagger \phi_3|^2 + |\phi_2^\dagger \phi_3 - \phi_3^\dagger \phi_1|^2 + |\phi_3^\dagger \phi_1 - 
\phi_1^\dagger \phi_2|^2\right)\,.
\label{VZ3Z3D8b}
\eea
Its symmetry group $\Sigma(36) = (\mathbb{Z}_3\times \mathbb{Z}_3)\rtimes \mathbb{Z}_4$ 
is generated by arbitrary permutations of the three doublets, by the discrete phase rotations
diag$(1,\omega,\omega^2)$, where $\omega = \exp(2\pi i/3)$, and by the following transformation
of order four:
\be
d = {1 \over \omega^2-\omega} \left(\begin{array}{ccc} 1 & 1 & 1 \\ 1 & \omega^2 & \omega \\ 1 & \omega & \omega^2 \end{array}\right)\,.
\ee
It is remarkable that this potential has only one ``structural'' free parameter $\lambda$
and that the term containing it reduces the full $PSU(3)$ symmetry group to a finite subgroup.

\subsection{Including generalized-$CP$ transformations}

One of the motivations of using more than one Higgs doublet is the possibility to generate 
spontaneous $CP$-violation. In 2HDM it was suggested by T.D.~Lee \cite{TDLee} back in 1973,
while with three doublets it is possible to make the relative phases of vacuum expectation values
stable against perturbations (so-called geometric phases). 
It is therefore interesting to see when the model becomes explicitly $CP$-conserving
and how this fact can be related with the Higgs-family symmetry group.
For example, it is known that in 2HDM {\em any} Higgs-family symmetry automatically entails
explicit $CP$-conservation.

We explored these issues in \cite{abelianNHDM} for abelian groups 
and in \cite{finite3HDMlong} for non-abelian finite groups in 3HDM.
For each group in the list (\ref{list}), we found conditions on the parameters of the potential 
when the potential becomes invariant under a generalized-$CP$ transformation
(but preserving its Higgs-family symmetry group). The corresponding list of groups
can be found in \cite{finite3HDMlong}. Here we quote only three results.

First, unlike 2HDM, it is possible to construct an explicitly $CP$-violating 3HDM potential 
with a non-trivial Higgs-family symmetry. This shows that the relation between the explicit 
$CP$-conservation and a Higgs-family symmetry is not as straightforward as in 2HDM.
Second, it turns out that a {\em sufficiently high} Higgs-family symmetry nevertheless
enforces explicit $CP$-conservation (for example, (\ref{VZ3Z3D8b}) is obviously invariant under
$CP$-transformation). Third, we found that presence of the $\mathbb{Z}_4$
symmetry group is a sufficient condition for explicit $CP$-conservation.

\section*{Acknowledgments}

This work was supported by the Belgian Fund F.R.S.-FNRS, 
and in part by grants RFBR 11-02-00242-a, RFBR 10-01-0391, RF President grant for
scientific schools NSc-3802.2012.2, and the 
Program of Department of Physics SC RAS and SB RAS "Studies of Higgs boson and exotic particles at LHC."

\end{document}